\documentclass[conference]{IEEEtran}
\ifCLASSINFOpdf
   \usepackage[pdftex]{graphicx}
\else
   \usepackage[dvips]{graphicx}
\fi
\usepackage[cmex10]{amsmath}
\usepackage{mathtools}
\usepackage{amsfonts}

\begin{document}
%
\title{Incorporating Spontaneous Reporting System Data to Aid Causal Inference in Longitudinal Healthcare Data}

\author{\IEEEauthorblockN{Jenna M. Reps and Uwe Aickelin}
\IEEEauthorblockA{School of Computer Science\\
University of Nottingham\\
Nottingham, NG8 1BB\\
Email: \{jenna.reps, uwe.aickelin\}@nottingham.ac.uk}
}


%


\maketitle

\begin{abstract}
Inferring causality using longitudinal observational databases is challenging due to the passive way the data are collected.  The majority of associations found within longitudinal observational data are often non-causal and occur due to confounding. 

The focus of this paper is to investigate incorporating information from additional databases to complement the longitudinal observational database analysis.  We investigate the detection of prescription drug side effects as this is an example of a causal relationship.  In previous work a framework was proposed for detecting side effects only using longitudinal data. In this paper we combine a measure of association derived from mining a spontaneous reporting system database to previously proposed analysis that extracts domain expertise features for causal analysis of a UK general practice longitudinal database.

The results show that there is a significant improvement to the performance of detecting prescription drug side effects when the longitudinal observation data analysis is complemented by incorporating additional drug safety sources into the framework.   The area under the receiver operating characteristic curve (AUC) for correctly classifying a side effect when other data were considered was 0.967, whereas without it the AUC was 0.923 However, the results of this paper may be biased by the evaluation and future work should overcome this by developing an unbiased reference set.


\end{abstract}


%
\IEEEpeerreviewmaketitle

\section{Introduction}
The current gold standard methodology for inferring causality between drugs and health outcomes is to conduct a randomised clinical trial \cite{cochran1973controlling}.  Methods have been developed for identifying associations between drugs and health outcomes using longitudinal observational data but due to the passive way that data are collected, confounding is a common occurrence \cite{Reps2013}.  Confounding is when an association between two variables is identified but the association is caused by a third unobserved variable being associated to both of the variables.  Due to the problem of confounding, relationships between drugs and health outcomes that are detected in longitudinal observational databases often require further analysis before causality is confirmed.  This additional analysis is often in the form of experimentation via randomised trails.  This is costly, sometimes unethical and cannot always be implemented \cite{black1996we}.  This issue has motivated an active field of research into methods that can identify causal relationships without requiring additional experimentation.

In previously work, researchers have investigated using more advanced supervised data mining methods to identify causality in longitudinal observational databases.  Examples include creating constrained Bayesian networks \cite{cooper1997simple} or creating features based on domain expertise in causal inference \cite{Repsthesis}.  In the later work, the authors proposed generating attributes based on the nine Bradford Hill causality considerations \cite{Hill1965} that are often used by epidemiologists when manually determining causality between drugs and health outcomes.  Training a classifier to distinguish between causal and non-causal relationships using five of the Bradford Hill causality consideration proposed attributes lead to a lower false positive rate that previously obtained using unsupervised methods  \cite{Repsthesis} and was suitable for causal inference with big data.  Unfortunately the false positive rate was still higher than desired, motivating further development of the idea by incorporating more of the Bradford Hill causality considerations.   In this paper we investigate incorporating the consistency consideration from the Bradford Hill causality considerations and determine whether adding this consideration improves the classification.

The consistency consideration referred to whether an association is found consistently across diverse and disperse sources of data.  If a drug truly causes a specific health outcome, then the association between the drug and health outcome should be found in different sources of data.  When an association is only found in one data source, then there is a good chance that it may just have occurred by chance or due to some form of bias in that way the data were collected. To incorporate the consistency consideration into the causal inference model perviously developed we calculate a measure of association using the USA's Food and Drug Administrations Adverse Event Reporting System (FAERS) data \cite{ahmad2003adverse} to complement the analysis applied to a UK general practice database known as The Health Improvement Network (THIN) database (www.thin-uk.com) \cite{THIN}.

The continuation of this paper is as follows.  In section \ref{back} we discuss the importance of incorporating expert domain knowledge for successful data mining and describe the existing causal inference method based on the Bradford Hill considerations. In section \ref{mat} we describe the data used throughout this paper and the various measures used to evaluate the causal inference method.  This is followed by the new framework that incorporates the consistently consideration in section \ref{frame}.  In section \ref{dis} we present the results of the analysis on a reference set and discuss these results.  The paper concluded with section \ref{conc}.

\section{Background} \label{back}
There is debate about whether it is domain expertise or machine learning skills that are the most important factor for successful data mining.  It is a generally accepted that domain expertise is important in all aspects of the knowledge discovery process \cite{fayyad1996kdd}.  Making use of domain expertise to understand the problem enables the data miner to extract suitable features and pre-process the data in a way that enables classifiers to distinguish between classes.  With well-designed and relevant features, it is possible that the classes are separable in the feature space.  In this situation, any classifier should perform reasonably well.  However, if the features are unsuitable then the majority of classifiers will perform poorly and advanced techniques are required.  Therefore, whenever possible, it is important to incorporate domain expertise into the feature extraction to simplify the classification task. 

In \cite{Repsthesis} the authors incorporate causal inference domain expertise to extract features that could be used as input into training a classifier to identifying causal relationships between drugs and health outcomes.  The features were extracted based on Bradford Hill's causality considerations \cite{Hill1965}.  These are a set of nine considerations that are often used to identify a causal relationship such as a drug's side effects.  The considerations are:

\begin{description}
\item[i)] Association strength: A measure of dependancy between the drug and health outcome.
\item[ii)] Temporality: Does the drug occur before the health outcome or the health outcome before the drug?
\item[iii)] Specificity: Is the drug only associated to one health outcome and the health outcome only associated to one drug?
\item[iv)] Consistency: Is there evidence of the association in difference sources of data?
\item[v)] Biological gradient: Is there a correlation between the dosage of the drug and the occurrence of the health outcome?
\item[vi)] Experimentation: Does stoping the drug stop the health outcome and restarting the drug restart the health outcome?
\item[vii)] Coherence: Does the drug causing the health outcome make sense or would it contradict known knowledge?
\item[viii)] Plausibility: Is the health outcome a possible side effect of the drug (e.g. is there knowledge that the chemical structure may interact with some biological pathway to cause the health outcome)?
\item[ix)] Analogy: Is a similar drug know to cause the health outcome or the drug known to cause a similar health outcome?
\end{description}

In previous work, the classifier was trained to predict whether a drug and health outcome pair correspond to an adverse drug reaction based on the extraction of their features from the longitudinal observational data.  The extracted features corresponded to the drug and health outcome relationship's association strength, temporality, specificity, biological gradient and experimentation.  This framework considering these five Bradford Hill considerations resulted in AUC values ranging between 0.883-937 \cite{Repsthesis}.  The analogy consideration was not used to create features, but was indirectly incorporated by applying a supervised learning technique.  The knowledge of drug and health outcomes that are known to correspond to adverse drug reactions or non-adverse drug reactions are utilised by the classifier to enable it to learn to predict whether a drug and health outcome pair correspond to an adverse drug reaction based on their extracted Bradford Hill derived features.

The classifier performed well and it was shown that including features based on Bradford Hill's specificity, biological gradient and experimentation considerations rather than just association strength and temporality significantly improved the ability to identify adverse drug reactions.  Unfortunately, due to restricting the analysis to a single database in previous work, it was not possible to extract  features based on the consistency consideration.  The plausibility and coherence considerations were also not previously used as these require expert knowledge about the chemical structure of the drug and known biological pathway interactions.  However, the plausibility and coherence considerations could be included in future work by incorporating chemical structure data. 

In this work we propose a way of combining the spontaneous reporting system databases with the longitudinal observational database analysis and can therefore create features corresponding to the consistency consideration.  It is of interest to determine whether including a different data source can improve the framework's adverse drug reaction detecting performance.  The FAERS database is partitioned by year and quarter. It would be possible to extract a measure of association for each drug and health outcome within the FAERS for each year from 2010 to 2013.  A drug and health outcome with a strong association in the THIN data and an association that occurs frequently across the FAERS records would be evidence of the drug and health outcome corresponding to an adverse drug reaction.

\section{Materials} \label{mat}
\subsection{THIN}    
\begin{figure}[tn]
\caption{A screen shot of the THIN therapy table}
\label{ther}
\includegraphics[width=0.5\textwidth]{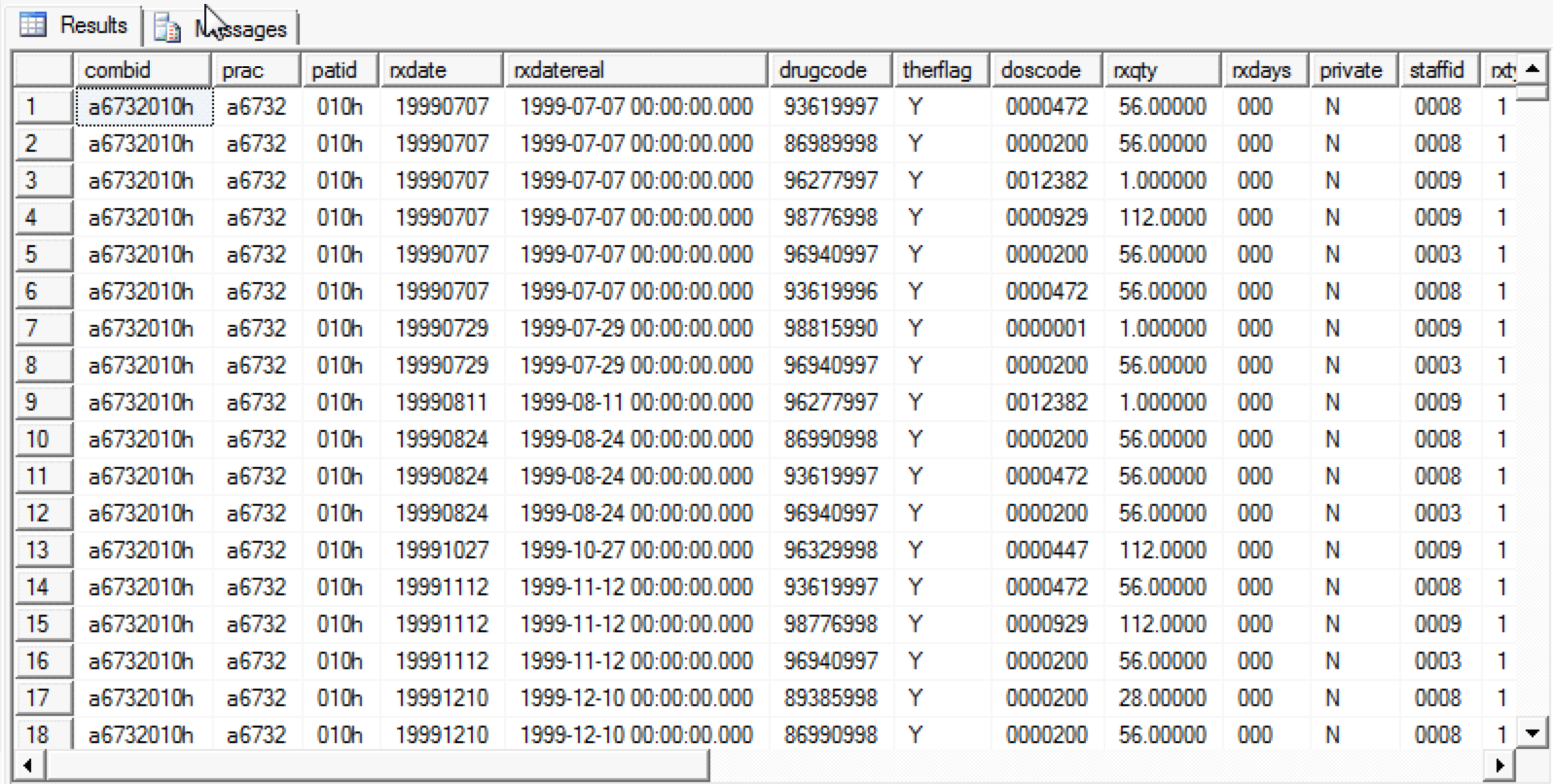}
\end{figure}

\begin{figure}[tn]
\caption{A screen shot of the THIN patient table}
\label{pat}
\includegraphics[width=0.5\textwidth]{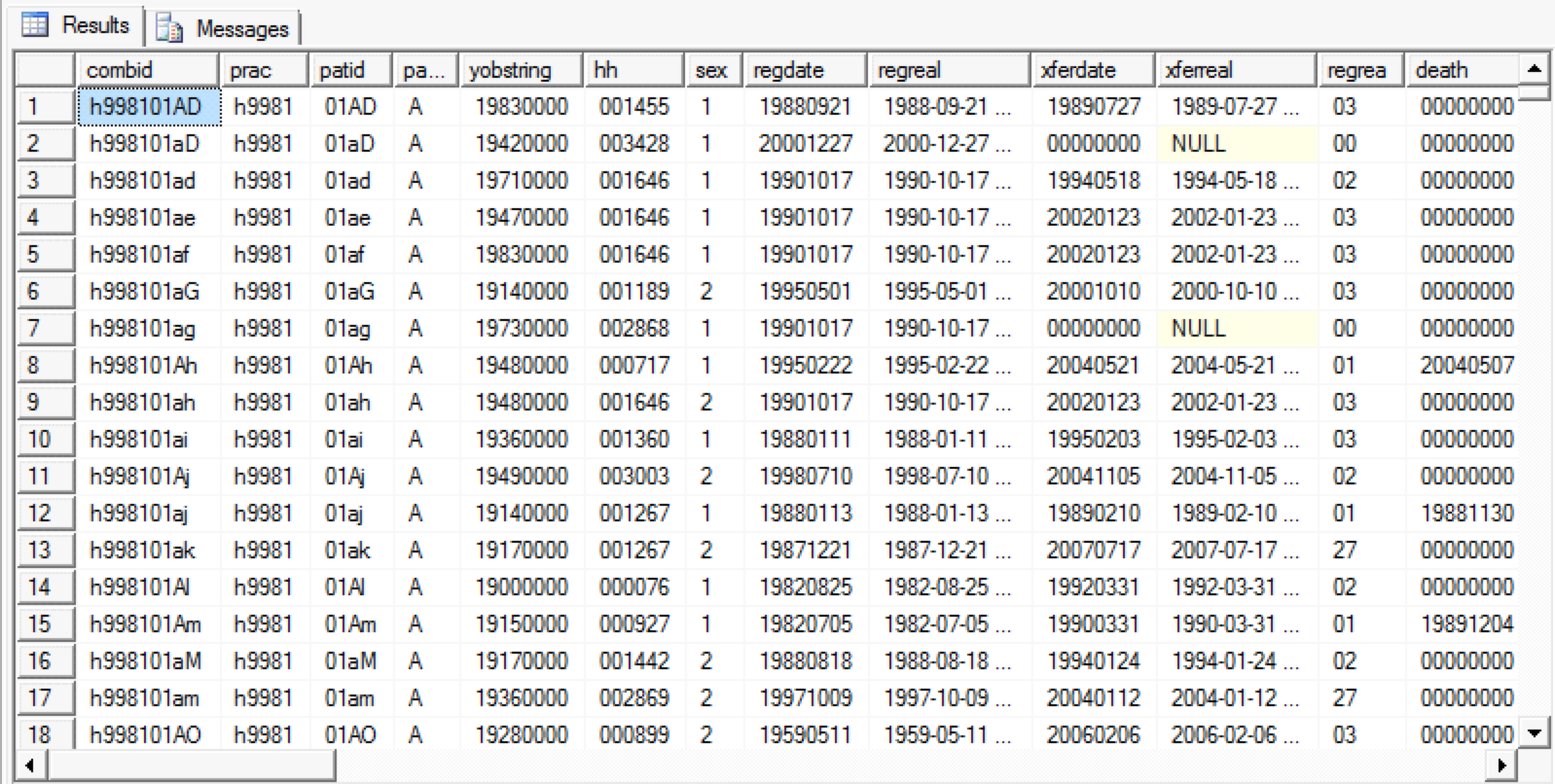}
\end{figure}

\begin{figure}[tn]
\caption{A screen shot of the THIN medical table}
\label{med}
\includegraphics[width=0.5\textwidth]{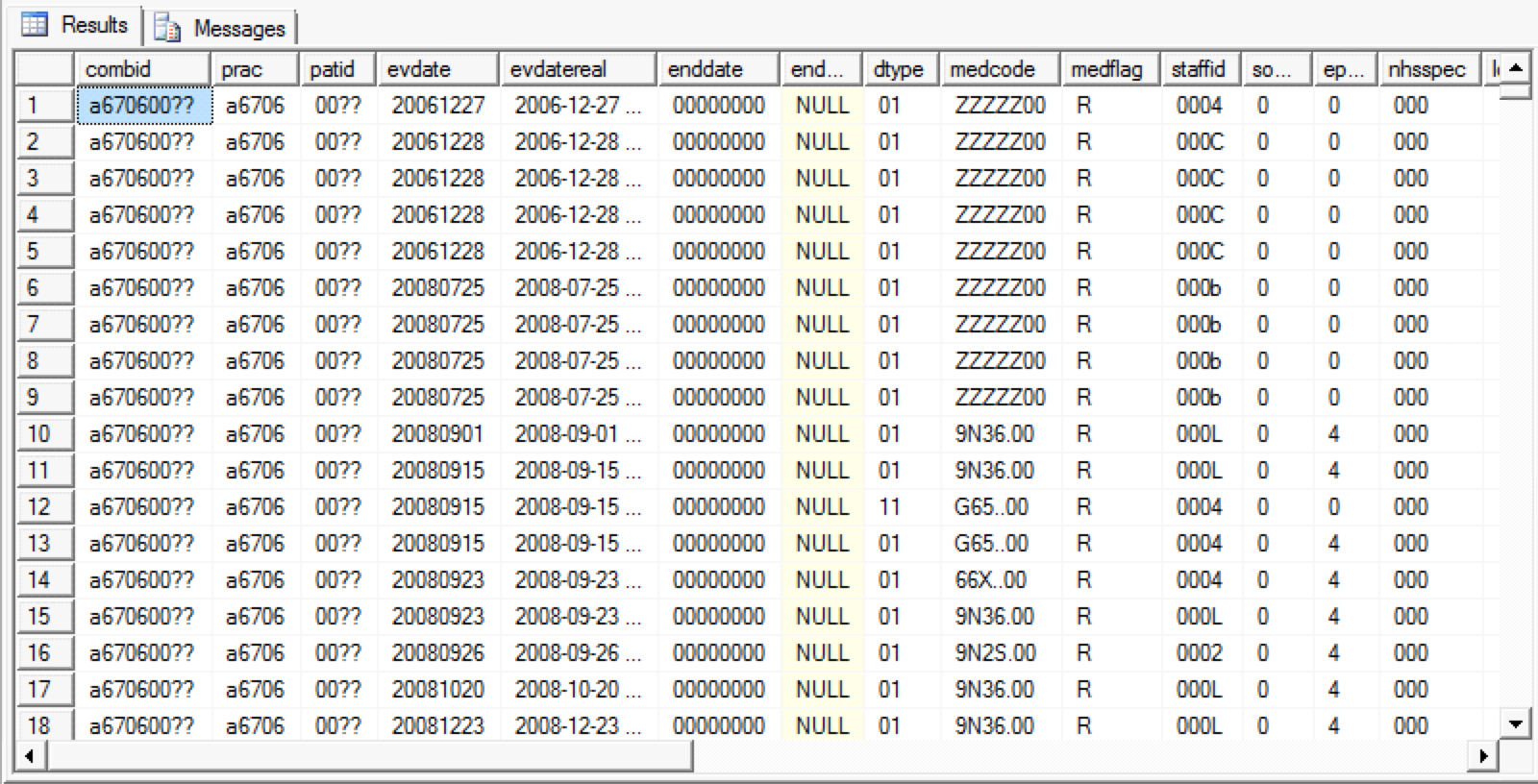}
\end{figure}

The THIN database is a longitudinal observational database containing general practice data from the UK.  The data are extracted directly from the local databases of the 587 participating general practices and are then validated and anonymised.  The complete database contains over 3.6 million active patient and over 12 million patients in total.  For each patient their year of birth and gender are recorded.  There is also additional demographic data often recorded.  While patients are registered at the general practice and it is participating, any medical events (e.g., diagnosis, symptom, laboratory test or administration event) that the patient informs the general practice of is recorded into a medical table with a corresponding date of recording.  Any drugs that are prescribed during this period are recorded into a therapy table along with the date of the prescription.  The THIN database contains over 750 million medical records and over 1 billion therapy records. Screen shots of the therapy, patient and medical tables contained in THIN are displayed in Fig. \ref{ther} - Fig. \ref{med}.
\begin{figure}[tn]
\caption{An example of the hierarchical structure of the READ codes}
\label{readcode}
\includegraphics[width=0.5\textwidth]{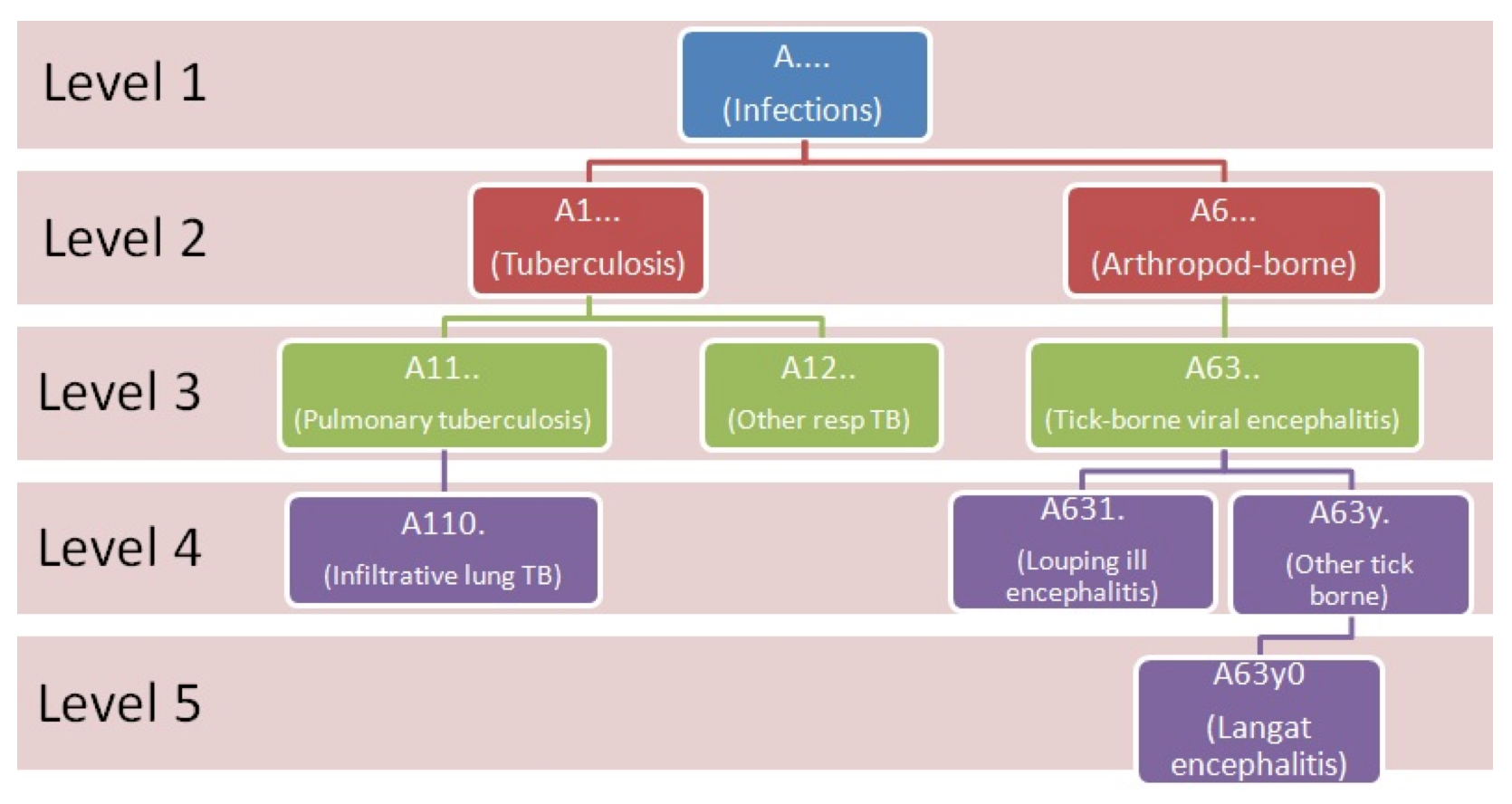}
\end{figure}

The medical events are recorded via a clinical encoding consisting of 5 alphanumerics/dot characters known as a READ code \cite{chisholm1990read}. Each READ code is linked to a description string detailing the medical event. The level of a READ code $x=x_{1}x_{2}x_{3}x_{4}x_{5}$ is defined as $L(x)=max \{i : x_{i} \ne . \}$. The READ codes have a hierarchal structure with child READ codes corresponding to the same medical event as their parents but with more detail, see Fig. \ref{readcode}. A READ code,  $x=x_{1}x_{2}x_{3}x_{4}x_{5}$, is the parent of another READ code,  $y=y_{1}y_{2}y_{3}y_{4}y_{5}$ if the level of READ code  $x$ is one less then the level of READ code $y$ and $x_{i}=y_{i}, \forall i\in \mathbb{N} \le L(x)$.  For example, the READ code `A....' corresponds to the description `Infection' and is the parent of the READ code `A1...' corresponding to `Tuberculosis', which is the parent of the READ code  `A11..' corresponding to `Pulmonary tuberculosis'. The drug prescriptions are recorded into the THIN database via a multilexeid code.  The multilexeid code has a corresponding string detailing the drug's generic name and dosage.  

In this paper we use a subset of the THIN database.  The subset consists of approximately half of the patients within the whole database but contains the complete medical and therapy records for these patients.  A subset of the THIN database is used in this research as this enables us to develop novel analytical techniques that will later be evaluated on the remaining THIN data.  The potential adverse drug reactions identified during the research on the first half of the THIN database can be evaluated with standard epidemiological analysis on the second half of the database.

There are some issues with the THIN database that can bias analysis.  One known problem is that patients can register at a new general practice at any point in time.  This can cause issues with the recording of their medical events, as it is common for newly registered patients to inform their new doctor of existing illnesses.  Due to them being at a new practice, the doctor will record these existing illnesses but the date will be the date they informed the doctor of these illnesses rather than the date that the illness first occurred.  Previous research has shown that the probability of patients informing their doctors of existing illnesses is reduced after being at the practice for 12 months \cite{lewis2005relationship}.  Therefore, we ignore the first 12 months of data for a newly registered patient.

\subsection{FAERS}
\begin{figure}[tn]
\caption{The structure of the old FAERS database from \cite{poluzzi2012data}.  The ISR has now been replaced by the primaryid and caseid}
\label{faers}
\includegraphics[width=0.5\textwidth]{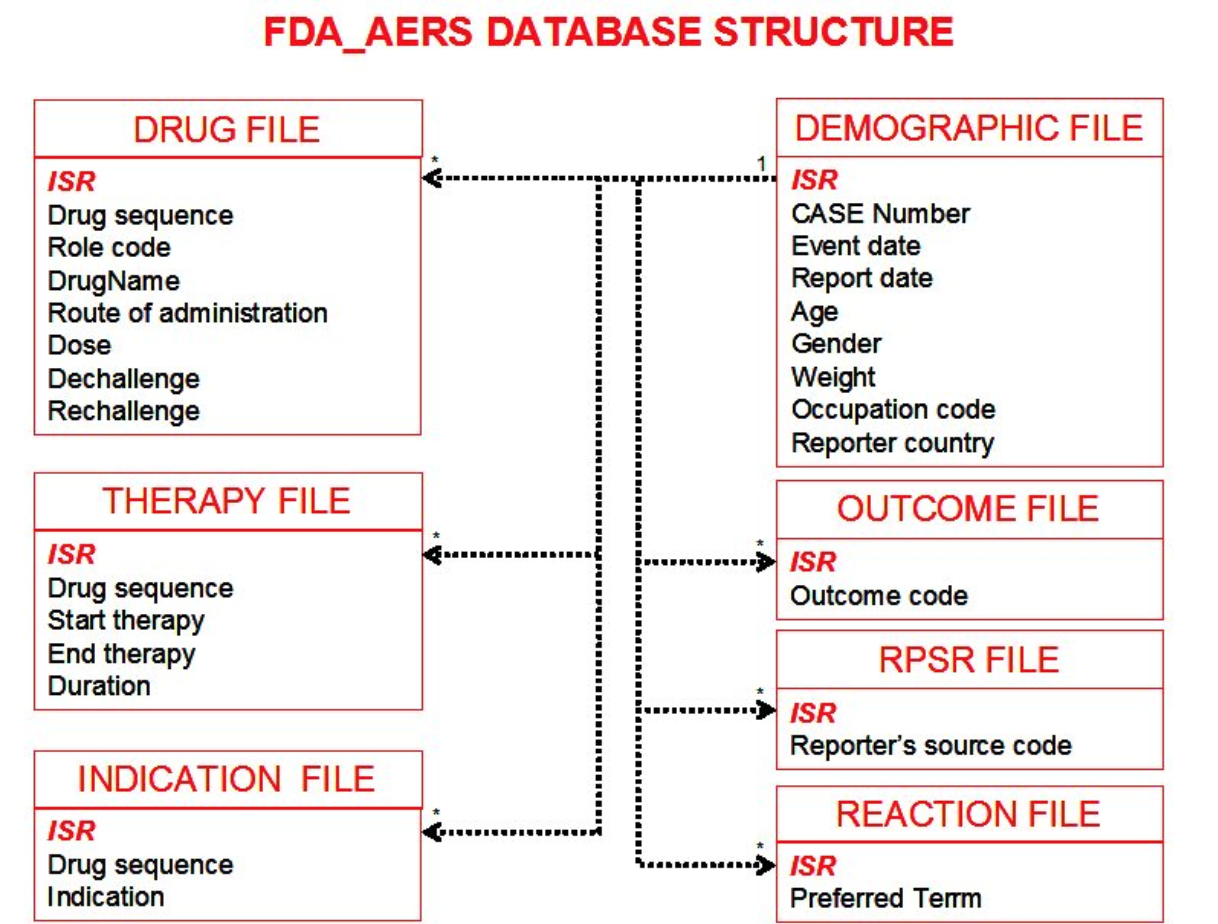}
\end{figure}

The FAERS is a spontaneous reporting system (SRS) database collect in the USA, see Fig. \ref{faers} for the database structure of the FAERS.  SRS databases contain records of suspected adverse drug reactions.  Medical health practitioners or the consumers, such as patients, can submit a record in a spontaneous reporting system if they expect they have witnessed or experienced an adverse drug reaction.  The records therefore contain a link between a drug or set of drugs and a medical event.  The data are stored for each year and quarter. In this paper we used the FAERS data from 2010 Q1 - 2013 Q4.  We combined Q1-Q4 reports each year, so we had four datasets, the reports recorded in years 2010, 2011, 2012 and 2013.

The FAERS data contain seven tables:
\begin{itemize}
 \item Therapy- contains the start and end day of the prescription
\item Drug - contains drug name and dosage information
\item Reaction - contains the suspected adverse event
\item Outcome - contains the outcome of the suspected adverse drug reaction
\item Demographics - contains details about the patient
\item Indication - contains the cause of the patient taking the prescription   
\item RPSR - contains information about the person submitting the report
\end{itemize}

The drug table contains details of the drug suspected to have caused an adverse drug reaction.  The details include the drug's generic name in upper case, the drug dosage information and the role of the drug within the report (e.g. is it a primary suspect or concomitant).  The health outcome suspected to have been caused by an adverse drug reaction is recorded into the reaction table.  The column ISR, corresponding to independent safety report, historically linked the drug and reaction table records, however, in more recent files this has been replaced by caseid and primaryid.  Within the reaction table, the health outcome is recorded via a string detailing the health outcome.  The string comes from a coding system known as the Medical Dictionary for Regulatory Activities (MedDRA) \cite{brown1999medical}.  This coding system was developed specifically for drug safety purposes.  

As the THIN and FAERS have different recording codes for the medical events and drug prescriptions we will combine the records using string matching as both databases contain the medical event descriptions and generic drug name strings.

\subsection{SIDER}
The Bradford Hill based framework for discovering adverse drug reactions requires training a classifier to distinguish between adverse drug reactions and non-adverse drug reactions.  To train such a classifier requires a training set of labelled data.  This means we need to know a set of drug and health outcome pairs where the drug is known to cause the health outcome and a set of drug and health outcome pairs where the drug is known to not cause the health outcome.

To find a set of drug and health outcomes where the drug is known to cause the health outcome we used the online side effect resource known as SIDER \cite{SIDER}. SIDER contains drug and health outcome classifications.  A search can be implemented to find the set of health outcomes that are indications to a specific drug or known side effects. The authors used text mining to extract the drug packaging labelled adverse drug reactions and indications in addition to extracting information from public documents.  SIDER uses the medDRA coding system. 

\subsection{Non adverse events} \label{non}
To find a set of drug and health outcome pairs where the drug does not cause the health outcome we identified health outcomes that do not correspond to an actual illness or cannot be caused by a drug acutely.  This was accomplished due to the hierarchal nature of the READ codes.  We found parent READ codes such as `family history' or `cancer' or `history of' and selected all the child, grandchild or great grandchild READ codes.  These READ codes were considered not possible to be an acute adverse event. Any drug and READ code pair where the READ code was from the set of Non adverse events was deemed impossible to correspond to adverse drug reaction and could therefore be classed as a non-adverse drug reaction.

\subsection{Combining the Data Sources}
The SIDER and FAERS data are readily combined as they use the medDRA coding system.  Combining the THIN database presents a challenge as the medical events are recorded via the READ code system.  In this work we combined THIN, FAERS and SIDER by exact non-case sensitive string matching.  For each of the READ codes in THIN the corresponding description was matching with the medDRA code description.  For example, if in THIN the READ code's description was 'Vomiting', then we matched this record with any SIDER and FAERS record with a medDRA code description of 'vomiting'.  This may result in many unmatched THIN and FAERS/SIDER records that actually correspond to the same health outcome but have non-generic descriptions so the string descriptions are not exactly the same.

\subsection{Software}
The software used in this study was SQL to store and pre-process the data and the open software R \cite{R} to perform the analysis.  The classification was performed using the `caret' library \cite{kuhn2008building} and the evaluation was performed using the `pROC' library \cite{robin2011proc}. 

\section{Framework Incorporating Consistency} \label{frame}
\subsection{Data Creation}
The Bradford Hill framework requires extracting features from the THIN and FAERS databases for a collection of drug and health outcome pairs that are known to correspond to adverse drug reactions (using SIDER) or cannot correspond to an adverse drug reaction (due to selecting health outcome having a clear non-drug cause).

\subsubsection{Finding the labels}
The first step is to find the drug and health outcome pairs where there seems to be a temporal association between the drug and health outcome in THIN and a true label is known.  Given a selection of drugs, for each drug all the records of patients being prescribed the drug for the first time are extracted.  A drug and READ code pair is created for each READ code that was recorded within a month of the first prescription of the drug for three or more patients. The set containing all these pairs is $P=\{ p_{i} \}$.  For a drug and READ code pair $p_{i} \in P$, we then calculate the number of prescriptions of the drug where the READ code occurred in the month before the drug, $B_{i}$, and the number of prescriptions of the drug where the READ code occurred in the month after the drug, $A_{i}$.  All the drug and READ code pairs where the READ code occurred more often before the prescription were excluded, $\hat{P}=\{ p_{i} \in P : A_{i}/B_{i}>1 \}$.  The remaining drug and READ code pairs are the ones that appear to have an association in THIN.

Where possible these pairs are then labelled as corresponding to a known adverse drug reaction or non-adverse drug reaction.  This was accomplished by labelling any pair with a READ code from the non adverse events set detailed in section \ref{non} as a non-adverse drug reaction.  For the remaining unlabelled pairs, the READ code's description was matched with the known SIDER listed adverse drug reactions of the drug and any pair with a match was labelled as a known adverse drug reaction.  The unlabelled pairs were discarded. Formally, the label for $p_{i}\in \hat{P}$ is
\begin{equation}
y_{i} = \left\{ 
  \begin{array}{l l}
    1 & \quad \text{if $p_{i}$ is a known side effect on SIDER} \\
    0 & \quad \text{if the READ code of $p_{i}$ is } \\
      & \quad \text{not a possible adverse event } \\
   -1 & \quad \text{the label is unknown}
  \end{array} \right.
\end{equation}
the drug and READ code pairs of interest are then, $\bar{P}=\{ p_{i}\in \hat{P} : y_{i} \ge 0 \}$.  This resulted in a set of 8158 labelled drug and READ code pairs, with 733 labelled as known adverse drug reactions and 7425 labelled as non-adverse drug reactions.

\subsubsection{Extracting THIN features}
For a labelled drug and READ code pair, $p_{i}$, we extracted the association strength, temporality, specificity, experimentation and biological gradient features from the THIN database.  The extracted association strength features used various measures of risk.  The risk of a READ code during a defined time period for a set of patients is simply the number of patients who experience the READ code during the define time period divided by the number of patients.  The risk difference is the risk of the READ code during the month after the prescription for the one set of patients minus the risk of the READ code during the month after the prescription for a different set of patients.  The risk ratio  is the risk of the READ code during the month after the prescription for the one set of patients divided by the risk of the READ code during the month after the prescription for a different set of patients.  The odds ratio is odd of the READ code occurring during the month after the prescription for the one set of patients divided by the odds of the READ code occurring during the month after the prescription for a different set of patients.  The extracted features for the drug and READ code pair $p_{i}$ are; 
\begin{description}
\item[$x_{1}$:] The risk difference comparing the patients prescribed the drug and prescribed any other drug.
\item[$x_{2}$:] The risk ratio comparing the patients prescribed the drug and prescribed any other drug.
\item[$x_{3}$:] The odds ratio comparing the patients prescribed the drug and prescribed any other drug.
\item[$x_{4}$:] The risk difference comparing the patients prescribed the drug and prescribed any other drug but with an additional prescription filter.  The filter removed prescriptions from the THIN data of any drug where a drug from the same family was prescribed in the previous 12 months.  The risk difference was then calculated on the filtered THIN data. 
\end{description}

The temporality feature, $x_{5}$, is $A_{i}/B_{i}$.  The specificity features are:
\begin{description}
\item[$x_{6}$:] the average age of the patients prescribed the drug who have the READ code recorded within a month of the prescription divided by the average age of the patients prescribed the drug.
\item[$x_{7}$:] the gender ratio (males/females) of the patients prescribed the drug who have the READ code recorded within a month of the prescription divided by the gender ratio of the patients prescribed the drug.
\item[$x_{8}$:] the READ code level (L($p_{i}'s$ corresponding READ code).
\end{description}
The biological feature, $x_{9}$ ,  is the average drug dosage only considering the patients prescribed the drug who have the READ code recorded within a month of the prescription divided by the average drug dosage when considering all the patients prescribed the drug.  The experimentation feature, $x_{10}$ , calculates how many patients experience the READ code within a month after a prescription of the drug and not during the month before for two or more distinct prescriptions of the drug divided by the number of patients who have a distinct repeat prescription of the drug.

\subsubsection{Extracting consistency feature}
\begin{table}[t]
\centering
\caption{The contingency table often used for analysing SRS data such as FAERS.}
\label{cont}
\begin{tabular}{c|cc}
 & Health outcome  m & Other Health outcome \\ \hline
Drug n & a  & b \\
Other Drug & c & d \\
\end{tabular}
\end{table}

To extract features corresponding to the consistency consideration we calculated the measure of association between a drug and health outcome for each year of FAERS data.  The risk difference was used to determine a measure of association for each year of FAERS data, using the values in a Contingency table, see Table \ref{cont}. The risk difference calculation for drug n and health outcome m is
\begin{equation}
RD_{mn}=  [a/(a+b)]-[c/(c+d)]
\end{equation}

The consistency feature, $x_{11}$ , was then calculated as the number yearly FAERS datasets where the drug and health outcome had a positive risk difference.  For example, if the risk difference for a specific drug and health outcome was 0.4 when considering the 2010 FAERS data, 0.1 for the 2011 FAERS data, -0.05 for the 2012 FAERS data and the health outcome was not recorded with the drug in 2013, then $x_{11}=2$.

To combine the consistency feature for a drug and health outcome coded in medDRA with the THIN features we matched the READ code's description string with the FAERS's medDRA description string and the drug strings in THIN and FAERS. Table \ref{matching} illustrates the matching implemented.
\begin{table*}[t]
\centering
\caption{The THIN and FAERS data were combined when the outcomes and drugs matched exactly.}
\label{matching}
\begin{tabular}{c|cccc}
THIN Outcome & THIN Drug & FAERS Outcome & FAERS Drug & Match  \\ \hline
Nausea & Ciprofloxacin & NAUSEA & Ciprofloxacin & Yes \\
CO Nausea & Ciprofloxacin & NAUSEA & Ciprofloxacin & No \\
HO Nausea & Ciprofloxacin & Nausea & Ciprofloxacin & No \\
Nausea & Ciprofloxacin & NAUSEA & Cipro& No \\
Nausea NED & Ciprofloxacin & NAUSEA & Ciprofloxacin & No \\
\end{tabular}
\end{table*}

\subsection{The complete data}
This resulted in a vector of features $\mathbf{x_{i}} \in \mathbb{R}^{11}$ for each labelled drug and READ code pair, $p_{i} \in \bar{P}$.  Therefore the labelled data corresponding to $p_{i} \in \bar{P}$ are $X= \{(\mathbf{x_{i}}, y_{i})\}$.  For the 23 drugs investigated there were 8158 drug-READ code pairs that could be labelled, with 733 labelled as an adverse drug reaction.

\subsection{Evaluation}
The Bradford Hill framework's classifier is evaluated by finding how often the classifier correctly classifies a drug and READ code pair as corresponding to an adverse drug reaction.  The labelled data set, $X=\{ (\mathbf{x_{i}}, y_{i}) \}$, was partitioned into 80\% training/testing $X_{T}$ and 20\% validation $X_{V}$.  The classifier is trained on $X_{T}$ using 10-fold cross validation to learn a function $f:\mathbb{R}^{10} \to \{0,1\}$ that maps a drug and READ code pair's Bradford Hill based extracted features into a class of adverse drug reaction or class of non-adverse drug reaction.

The trained classifier is then applied to the extracted features of each drug and READ code pairs in the validation set to predict their classes, $f(\mathbf{x_{i}}), (\mathbf{x_{i}}, y_{i}) \in X_{V}$ and the prediction is compared with the truth. The classification is,
\begin{itemize}
\item TP  when $f(\mathbf{ x_{i}}) = 1 $ and $ y_{i}=1$
\item TN  when $f(\mathbf{x_{i}}) = -1 $ and $ y_{i}=-1$
\item FP  when $f(\mathbf{x_{i}}) = 1 $ and $ y_{i}=-1$
\item FN  when $f(\mathbf{x_{i}}) = -1 $ and $ y_{i}=1$
\end{itemize}

The sensitivity and specificity of the classifier are,
\begin{equation*}
Sensitivity = TP/(TP+FN)
\end{equation*}
\begin{equation*}
Specificity = TN/(FP+TN)
\end{equation*}

The receiver operating characteristic, ROC, curve is then drawn by plotting the sensitivity against one minus the specificity.  A common measure of performance for a classifier is the area under the ROC curve (AUC) \cite{cortes2004}.  As we are interested in a classifier that can identify adverse drug reactions without incorrectly classifying many non-adverse drug reactions, we also calculate the partial AUC between the specificity values 0.8-1, denoted pAUC$_{[0.8,1]}$.  The AUC of two classifier can be compared using the Delong method \cite{delong1988comparing} and we use this technique to determine significant differences at a 5\% significance level. 

\section{Results \& Discussion} \label{dis}
\begin{figure}
\includegraphics[width=0.5\textwidth]{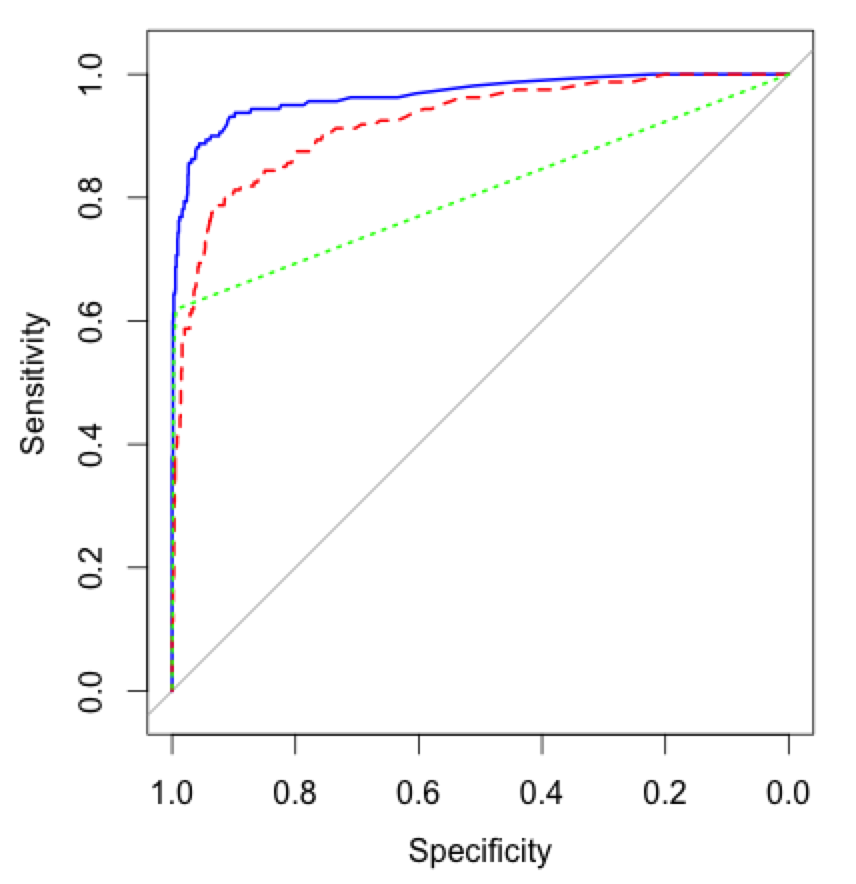}
\caption{The ROC plots for the Bradford Hill framework classifier not including the consistency feature (red), the Bradford Hill framework classifier including the consistency feature (blue) and the number of years that the FAERS data had a positive risk difference for the drug and READ code pair (green).}
\label{roc}
\end{figure}
\begin{table}
\caption{The AUC values for the different classifiers.}
\label{auc}
\begin{tabular}{c|cc}
Method & AUC & pAUC$_{[0.8,1]}$ \\
Framework incorporating the consistency feature & 0.967 & 0.1794 \\ \hline
Framework excluding the consistency feature & 0.923 & 0.1498 \\
The consistency feature alone & 0.807 & 0.1299  \\
\end{tabular}
\end{table}

The ROC plots for the Bradford Hill framework's classifier incorporating the consistency feature, the Bradford Hill framework's classifier excluding the consistency feature and just using the consistency feature are presented in Fig. \ref{roc}.  The AUC and pAUC$_{[0.8,1]}$ values are displayed in Table \ref{auc}.  It can be seen that incorporating the consistency feature significantly increased the AUC, 0.967 compared to 0.923 without the consistency feature (p-value 1.02 $\times 10^{-5}$). This shows that incorporating the consistency feature increased the frameworks ability to detect adverse drug reactions. This results also suggests that performing analysis by combining different sources of data can lead to improved results in health informatics.

The performance of just using the measure of consistency of an association between and drug and READ code pair over the years 2010-2014 within the FAERS data resulted in an AUC of 0.807.  The plot shows that the measure of consistency was able to identify many known adverse drug reactions, with a high sensitivity when the specificity is also high.  However, there is a point in the specificity where the measure of consistency is no longer able to identify adverse drug reactions.  This shows that the FAERS data can be used to identify adverse drug reactions accurately but is limited in that it cannot identify all the adverse drug reactions.  This highlights the requirement of performing analysis on the combination of longitudinal healthcare and SRS data to detect adverse drug reactions. 

The results suggest that the consistency feature extracted from the FAERS data is able to aid the classifier to detect adverse drug reactions that are not reported in the FAERS, as the framework incorporating the consistency feature outperformed the framework excluding the consistency feature and relying on the consistency alone.  We suspected that the inclusion of consistency feature may bias the classifier due to strong correlation between the number of positive risk difference values across the years 2010-2014 and the drug and READ code pair corresponding to an adverse drug reaction.  However, this was not the case, even though the consistency feature was highly skewed between the classes,  see table \ref{cons_dist}.  Over half of the ADRs could be identified, with a small false positive rate, using the signalling criteria of $x_{11} \geq 2$, however the THIN features were required to be able to signal the remaining ADRs that are reported less often in the FAERS data.
\begin{table}
\caption{Consistency attribute distribution across the classes.}
\label{cons_dist}
\begin{tabular}{c|ccccc}
            & $x_{11}=0$ & $x_{11}=1$ & $x_{11}=2$ & $x_{11}=3$ & $x_{11}=4$ \\ \hline
$y_{i}=0$ & 7391 & 15 & 6 & 5 & 8 \\
$y_{i}=1$ & 272 & 68 & 83 & 111 & 199\\
\end{tabular}
\end{table}

One limitation of this research is the potential bias of the data combination and labelling.  For example, the labels and consistency feature may highly correlated due to bias as the SIDER labels being derived from drug packaging and public documents that may have considered the SRS data.  The reference set of known non-adverse drug reactions also caused a bias as it is very difficult to know whether a health outcome is definitely not an adverse drug reaction to a specific drug. The reference set drug and health outcomes corresponding to non-adverse drug reactions were selected due to the health outcome having a clear non-drug cause.  Therefore the drug and health outcomes corresponding to non-adverse drug reactions in the reference set are extremely unlikely to be recorded as a suspected adverse drug reaction in the FAERS database. Both of these issue result in bias of the consistency attributes for the reference set used.  As a consequence the trained classifier is likely to predict any drug and health outcome pair that is recorded in FAERS, and therefore probably likely to have a consistency feature value greater than 0, as an adverse drug reaction.  However, many of the FAERS records may not correspond to an actual adverse drug reaction.  In future work it is important to improve the reference set by including drug and health outcome pairs that are known to correspond to non-adverse drug reaction but are still plausible (i.e., include health outcomes that are common illnesses such as `vomiting' or `rash').  Evaluating the framework on such a reference set will result in a less biased measure of how well the framework incorporating the Bradford Hill consistency consideration performs.

The framework was also limited by the string matching between the THIN READ code descriptions and the medDRA descriptions.  Many of the known adverse drug reactions may not be labelled in the data due to the READ code description slightly differing from the medDRA description and some of the drug and READ code pairs may have missing consistency feature values due to problems with the string matching.  If a natural language processing method was developed for mapping the READ code and medDRA description (or any medical terminology coding system) then it would enable different sources of data to be readily integrated and analysed together.  This is likely to help researchers extract new knowledge.

The framework incorporating the Bradford Hill association strength, temporality, consistency, specificity, biological gradient and experimentation has a high performance but this may be increased by including the plausibility and coherence considerations.  Other sources of data have been used to identify potential adverse drug reactions, including chemical structure data.  It may be possible to combine more sources of data, such as chemical structure databases, to cover all the Bradford Hill considerations and developed a framework that can detect any adverse drug reaction with an even higher specificity and sensitivity.

\section{Conclusion} \label{conc}
In this paper we have proposed a way to incorporate a measure of how consistent an association between and drug and health outcome is by combining different forms of drug safety data.  This increased the existing Bradford Hill based causal inference framework's ability to identify adverse drug reactions in longitudinal observational data.  The results show that incorporating features derived from the FAERS database significantly improved the classifiers ability to distinguish between adverse drug reaction relationships and non-adverse drug reaction relationships.

In future work a new reference set could be developed to evaluate the framework fairly.  It would also be of interest to incorporate chemical structure databases to include features based on the plausibility and coherence Bradford Hill considerations.





\bibliographystyle{IEEEtran}
\bibliography{refs}

\begin{thebibliography}{10}
\providecommand{\url}[1]{#1}
\csname url@samestyle\endcsname
\providecommand{\newblock}{\relax}
\providecommand{\bibinfo}[2]{#2}
\providecommand{\BIBentrySTDinterwordspacing}{\spaceskip=0pt\relax}
\providecommand{\BIBentryALTinterwordstretchfactor}{4}
\providecommand{\BIBentryALTinterwordspacing}{\spaceskip=\fontdimen2\font plus
\BIBentryALTinterwordstretchfactor\fontdimen3\font minus
  \fontdimen4\font\relax}
\providecommand{\BIBforeignlanguage}[2]{{%
\expandafter\ifx\csname l@#1\endcsname\relax
\typeout{** WARNING: IEEEtran.bst: No hyphenation pattern has been}%
\typeout{** loaded for the language `#1'. Using the pattern for}%
\typeout{** the default language instead.}%
\else
\language=\csname l@#1\endcsname
\fi
#2}}
\providecommand{\BIBdecl}{\relax}
\BIBdecl

\bibitem{cochran1973controlling}
W.~G. Cochran and D.~B. Rubin, ``Controlling bias in observational studies: A
  review,'' \emph{Sankhy{\=a}: The Indian Journal of Statistics, Series A}, pp.
  417--446, 1973.

\bibitem{Reps2013}
J.~M. Reps, J.~M. Garibaldi, U.~Aickelin, D.~Soria, J.~Gibson, and R.~Hubbard,
  ``Comparison of algorithms that detect drug side effects using electronic
  healthcare databases,'' \emph{Soft Computing}, vol.~17, no.~12, pp.
  2381--2397, 2013.

\bibitem{black1996we}
N.~Black, ``Why we need observational studies to evaluate the effectiveness of
  health care,'' \emph{British Medical Journal}, vol. 312, no. 7040, pp.
  1215--1218, 1996.

\bibitem{cooper1997simple}
G.~F. Cooper, ``A simple constraint-based algorithm for efficiently mining
  observational databases for causal relationships,'' \emph{Data Mining and
  Knowledge Discovery}, vol.~1, no.~2, pp. 203--224, 1997.

\bibitem{Repsthesis}
\BIBentryALTinterwordspacing
J.~M. Reps, ``Detecting adverse drug reactions in the general practice
  healthcare database,'' PhD Thesis, School of Computer Science, The University
  of Nottingham, 2014. [Online]. Available:
  \url{http://ima.ac.uk/wp-content/uploads/2014/08/thesis1.pdf}
\BIBentrySTDinterwordspacing

\bibitem{Hill1965}
A.~B. Hill, ``The environment and disease: association or causation?''
  \emph{Proceedings of the Royal Society of Medicine}, vol.~58, no.~5, p. 295,
  1965.

\bibitem{ahmad2003adverse}
S.~R. Ahmad, ``Adverse drug event monitoring at the {Food and Drug
  Administration},'' \emph{Journal of General Internal Medicine}, vol.~18,
  no.~1, pp. 57--60, 2003.

\bibitem{THIN}
J.~D. Lewis, R.~Schinnar, W.~B. Bilker, X.~Wang, and B.~L. Strom, ``Validation
  studies of the health improvement network ({THIN}) database for
  pharmacoepidemiology research,'' \emph{Pharmacoepidemiology and Drug Safety},
  vol.~16, no.~4, pp. 393--401, 2007.

\bibitem{fayyad1996kdd}
U.~Fayyad, G.~Piatetsky-Shapiro, and P.~Smyth, ``The {KDD} process for
  extracting useful knowledge from volumes of data,'' \emph{Communications of
  the {ACM}}, vol.~39, no.~11, pp. 27--34, 1996.

\bibitem{chisholm1990read}
J.~Chisholm, ``The {Read} clinical classification.'' \emph{British Medical
  Journal}, vol. 300, no. 6732, p. 1092, 1990.

\bibitem{lewis2005relationship}
J.~D. Lewis, W.~B. Bilker, R.~B. Weinstein, and B.~L. Strom, ``The relationship
  between time since registration and measured incidence rates in the general
  practice research database,'' \emph{Pharmacoepidemiology and Drug Safety},
  vol.~14, no.~7, pp. 443--451, 2005.

\bibitem{poluzzi2012data}
E.~Poluzzi, E.~Raschi, C.~Piccinni, and F.~De~Ponti, ``Data mining techniques
  in pharmacovigilance: analysis of the publicly accessible {FDA} adverse event
  reporting system ({AERS}),'' \emph{Data Mining Applications in Engineering
  and Medicine. Croatia: InTech}, pp. 267--301, 2012.

\bibitem{brown1999medical}
E.~G. Brown, L.~Wood, and S.~Wood, ``The medical dictionary for regulatory
  activities ({MedDRA}),'' \emph{Drug Safety}, vol.~20, no.~2, pp. 109--117,
  1999.

\bibitem{SIDER}
M.~Kuhn, M.~Campillos, I.~Letunic, L.~Jensen, and P.~Bork, ``A side effect
  resource to capture phenotypic effects of drugs,'' \emph{Molecular Systems
  Biology}, vol.~6, no. 343, 2010.

\bibitem{R}
\BIBentryALTinterwordspacing
{R Development Core Team}, \emph{R: A language and environment for statistical
  computing}, R Foundation for Statistical Computing, Vienna, Austria, 2008,
  {ISBN} 3-900051-07-0. [Online]. Available: \url{http://www.R-project.org}
\BIBentrySTDinterwordspacing

\bibitem{kuhn2008building}
M.~Kuhn, ``Building predictive models in {R} using the caret package,''
  \emph{Journal of Statistical Software}, vol.~28, no.~5, pp. 1--26, 2008.

\bibitem{robin2011proc}
X.~Robin, N.~Turck, A.~Hainard, N.~Tiberti, F.~Lisacek, J.-C. Sanchez, and
  M.~M{\"u}ller, ``{pROC}: an open-source package for {R} and {S}+ to analyze
  and compare {ROC} curves,'' \emph{BMC Bioinformatics}, vol.~12, no.~1, p.~77,
  2011.

\bibitem{cortes2004}
C.~Cortes and M.~Mohri, ``{AUC} optimization vs. error rate minimization,''
  \emph{Advances in Neural Information Processing Systems}, vol.~16, no.~16,
  pp. 313--320, 2004.

\bibitem{delong1988comparing}
E.~R. DeLong, D.~M. DeLong, and D.~L. Clarke-Pearson, ``Comparing the areas
  under two or more correlated receiver operating characteristic curves: a
  nonparametric approach,'' \emph{Biometrics}, pp. 837--845, 1988.

\end{thebibliography}
%

\end{document}